\documentclass[12pt]{article}
\usepackage{amsmath,amssymb,amsthm,latexsym}
\usepackage{stmaryrd,wasysym,upgreek,mathrsfs,dsfont}
\usepackage[english]{babel}
\usepackage{graphicx,color}
\usepackage{amscd}

\usepackage[pdftex]{hyperref}

\newcommand{\be}{\begin{equation}}
\newcommand{\bee}{\begin{equation}}
\newcommand{\ee}{\end{equation}}
\newcommand{\bea}{\begin{eqnarray}}
\newcommand{\eea}{\end{eqnarray}}

\begin{document}

\title{The Tensor Theory Space}


\author{Vincent Rivasseau\\
Laboratoire de Physique Th\'eorique, CNRS UMR 8627,\\
Universit\'e Paris XI,  F-91405 Orsay Cedex\\
and Perimeter Institute for Theoretical Physics,\\
31 Caroline St. N, N2L 2Y5, Waterloo, ON, Canada\\
E-mail: rivass@th.u-psud.fr}

\maketitle
\begin{abstract}

The tensor track is a background-independent
discretization of quantum gravity which includes a sum over all topologies. 
We discuss how to define a functional renormalization group flow and the
Wetterich equation in the corresponding \emph{theory space}. This 
space is different from the Einsteinian theory space of asymptotic safety. 
It includes all fixed-rank tensor-invariant interactions, hence generalizes
matrix models and the (Moyal) non-commutative field theory space.

\end{abstract}

\section{Introduction}

Quantum gravity on a flat fixed four-dimensional background is not perturbatively renormalizable \cite{Goroff:1985th}.
One can address the issue in many ways. The most conservative, asymptotic safety \cite{Weinberg,Reuter,ReuterSaueressig}, 
searches for a non-trivial ultraviolet fixed point in the ``Einsteinian" theory space of 
coordinate-invariant functions of the metric on a fixed background. Giving up Lorentz invariance
enlarges this theory space and allows to use more convergent propagators, like in solid state physics, leading to
perturbatively renormalizable theories \cite{Horava}. 
Other proposals such as loop quantum gravity \cite{Rovelli}, group field theory \cite{Oriti:2013aqa,
Krajewski:2012aw}Ê
and dynamical triangulations \cite{ambjorn-book} formulate the theory in a ``background independent"
way\footnote{Background independent schemes such as \emph{causal} dynamical triangulations may also break Lorentz invariance 
and use a preferred time \cite{Ambjorn:2013apa}.}. 
Remark that this requires to blur the ordinary notion of locality. In this case the effective geometry of our universe could 
emerge from more fundamental degrees of freedom, hence almost ``from scratch" \cite{scratch}. 
Supergravity keeps a fixed background but enlarges the symmetries of the theory to at least soften the ultraviolet divergencies. 
Superstring theory, the most developed and dominant current approach, combines supersymmetry with
extended objects (strings) moving on a fixed background, but its second quantization ($M$ theory) should be background independent.

The tensor track \cite{Rivasseau:2011hm,Rivasseau:2012yp,Rivasseau:2013uca} belongs to the background-invariant approaches
and started as an improvement of group field theory \cite{color} to allow for renormalization \cite{Rivasseau:2011xg}. It also 
generalizes renormalizable non-commutative field theory on Moyal space \cite{Grosse:2004yu}, which includes the very interesting features
of asymptotic safety \cite{Disertori:2006nq} and solvability in the planar sector \cite{Grosse:2012uv}.
It remains rather conservative in the sense that it sticks to quantum field theory and the renormalization group (RG),
but simply proposes a different \emph{theory space}. Up to now its most interesting result 
is the discovery that in this tensor theory space RG flows of perturbatively 
renormalizable models are generically asymptotically free \cite{BenGeloun:2012pu,BenGeloun:2012yk}. Hence random tensor theories could
form genuinely consistent ultraviolet completions of gravity. 

Asymptotic freedom seems a priori at odds
with the asymptotic safety paradigm. However, since the theory spaces are not the same, this contradiction
is only apparent. The two approaches may be not only compatible but ultimately complementary. 
Any interesting infrared phase in the tensor track will require 
a connection to the classical space-time we observe. This might require to weld at some point the 
tensor theory space and its renormalization group flow to the ones of asymptotic safety through a series of changes of variables. Indeed
we still hope that quantum gravity ought to be unique, or at least ``generic" in a certain sense. This idea may seem irrational at first sight.
But it has been already  fruitful, at least in dimension two. It led for instance to conjecture new relations in algebraic geometry \cite{witten}Ê
which quickly became theorems thanks to a matrix-model representation of these relations \cite{kontsevich}. In two dimensions, 
equivalence between the discrete matrix-based and the continuum Liouville-based formulations of quantum gravity 
is becoming progressively clearer \cite{davideynard}. Hopefully it will extend to higher dimensions as well, and ultimately
connect the discretized and continuum formulations of quantum gravity.

This paper is meant as a complement to the short review \cite{Rivasseau:2013uca}
which describes further aspects of the tensor track.

\section{FRG and the QEG theory space}

Renormalization group can be roughly described as a flow in a certain infinite dimensional functional space for actions, the theory space.
Scale plays the role of time for this flow.
Numerical analysis of non-trivial fixed points usually works within a finite dimensional truncation
of that theory space. In practice simple truncations usually detect quite easily non-trivial fixed points. For instance in the case of the discrete 
renormalization group flow for iterated maps of the interval, the theory space is made of (even) regular unimodal functions with a quadratic tip.
It has a non trivial fixed point governing the universal Feigenbaum exponents \cite{Feigenbaum:1977ys} for period doubling.
The existence of the fixed point can be proved rigorously \cite{lanford} and
truncations to (even) polynomials of rather small degree ($\le 12$) provide excellent approximations 
($\simeq 10^{-6}$) of the Feigenbaum exponents at the fixed point \cite{Gurau:2014vwa}.

In the usual field theory context the renormalization group, initially expressed by Wilson as a discrete step by step
evolution, can also be formulated in terms of differential equations for a continuous flow.
Let us describe briefly the functional renormalization group (FRG) based on Wetterich equation 
\cite{Wetterich:1992yh,Wetterich:1989xg} following \cite{Gurau:2014vwa}.

Consider the generating functional of the connected moments for a field $\phi$ with UV cut-off 
\begin{equation}
e^{W_\Lambda[J]}=\frac{1}{\mathcal{N}_\Lambda}\int d\mu_\Lambda ( \phi) \, e^{-S[\phi]+J\cdot\phi}\;,
\end{equation}
where the action includes the quadratic Gaussian part $S_{\rm free}$,
$d\mu_\Lambda ( \phi)$ is a functional measure with a ultraviolet cut-off at 
scale $\Lambda$, and the normalization is $\mathcal{N}_\Lambda=\int d\mu_\Lambda ( \phi)e^{-S_{\rm free}[\phi]}$.

Introducing an infrared cutoff which suppresses low energy modes ($E\lesssim k$) in the path 
integral, we associate to $W_{\Lambda}[J]$ a one parameter family of generating functionals
\begin{equation}\label{Wk}
e^{W_{k,\Lambda}[J]}=\frac{1}{\mathcal{N}_{k,\Lambda}}\int\ d\mu_\Lambda ( \phi)\, e^{-S[\phi]+J\cdot\phi-\Delta S_k[\phi]}\;  ,
\end{equation}
where now $\mathcal{N}_{k,\Lambda}=\int d\mu_\Lambda ( \phi) e^{-S_{\rm free}[\phi]-\Delta S_k[\phi]}$ and
\bee
\Delta S_k[\phi]=\!\int\! \frac{d^Dp}{2(2\pi)^D}\,\phi(p)\, \mathcal{R}_k(p)\,\phi(-p)  
\ee
regulates low-momentum modes, the regulator $\mathcal{R}_k(p)=  k^2\, r(p^2/k^2)$ being determined by the shape function $r(z)$. For instance the choice
$r(z)=\frac{z}{e^{z-1}}$ leads to a propagator with sharp infrared cutoff $k$ in parametric space
\bee  C_k   = \frac{1}{p^2} [ 1 - e^{-p^2/ k^2 }]   = \int_0^{k^{-2}} d\alpha e^{- \alpha p^2}.
\ee

Introducing the renormalization time $t=\log(k/\Lambda)$, so that $k\frac{d}{dk}=\frac{d}{dt}$,
and defining the Legendre transform $\tilde \Gamma_k$ of $W_k (J)$
leads to Wetterich equation for  the translated transform $\Gamma_k = \tilde \Gamma_k - \Delta S_k$
\bea\label{wetterich}
\frac{d \Gamma_k[\varphi]}{dt}&=&
\frac{1}{2} \mathrm{Tr}\left[\frac{\frac{d }{dt}\mathcal{R}_k}{\mathcal{R}_k+\Gamma^{(2)}[\varphi]}\right]\; ,
\eea
where $\Gamma^{(2)}= \frac{\delta^2\Gamma[\varphi]}{\delta\varphi\delta\varphi}$ and the trace stands for integration and summation on internal indices. 
The above functional renormalization group equation is formally 
a differential equation for the one-parameter family of functionals $\Gamma_k$. Its solution $\Gamma_k$ describes the flow of the effective
action in the theory space under changes of the cutoff scale. The similar Polchinski equation would compute the same flow 
but for connected functions. Hence the right-hand side of Polchinski's equation is made of two terms, 
one of which is one particle reducible. Wetterich equation gets rid of that term
by passing to one particle irreducible functions through the Legendre transform, 
hence it looks formally simpler. 

Numerical analysis of this equation 
projects $\Gamma_k$ onto a finite dimensional subspace $E_p$
of the theory space. This corresponds to restrict the theory to a finite set of $p$ coupling constants. 
The renormalization group flow is usually never stable under such restrictions, hence after each (discretized) elementary evolution step
of the flow, the evolved theory must be projected again onto the same subspace. Such a scheme clearly
does not reduce to a perturbative computation.

The asymptotic safety program, also named ``Quantum Einstein Gravity" (QEG),
uses as theory space the space of all regular functions of the metric $g_{\mu\nu}$ on a fixed ${\mathbb R}^4$ background 
with Euclidean or Lorentzian signature which are invariant under diffeomorphisms. 
This theory space is of course much more complicated than the single variable
functions for iteration of maps on the interval. A main difficulty is to carefully maintain
the diffeomorphisms symmetry which translate into Ward identities for the functional $\Gamma_k$ \cite{Reuter}.
The results are quite striking. A resilient ultraviolet fixed point with a generic characteristic 
shape has been found by all truncation schemes, starting with  the very simple Einstein-Hilbert truncation with cosmological constant, and including progressively
higher powers of the scalar curvature $R$ \cite{Codello:2007bd} which correspond to a local potential approximation \cite{Benedetti:2012dx},
then more recently the square of the Weyl curvature tensor \cite{Benedetti:2009rx}.

These results are both fascinating and frustrating. Ultimately the fixed point is the product of a numerical analysis
and there is not yet any simple explanation for its existence and properties. Also the formalism of QEG seems to exclude any space-time topology change. 
But excluding tunneling to other vacua typically leads to non unitarity in quantum field theory: probabilities no longer add up to one.
Finally an asymptotically safe trajectory does not describe any particular phase transition. This is a bit disappointing, given that nature seems to love them so much.
To rephrase this in a different way, nothing particularly special happens analytically
at the Planck scale in the asymptotic safety scenario, although convincing arguments point to quantum gravity implying a qualitative change in the physics
of space and time around that scale \cite{DFR}.
It is therefore tempting to consider the fixed point of QEG as an approximate origin for 
the flow of the quantum gravity action \emph{below} the Planck scale, rather than as the ultimate fundamental theory.
It would be nice to complete it with a more fundamental background independent theory 
whose own flow could provide, once the more fundamental variables are mapped
to a metric field on an emerging macroscopic space-time, some kind of effective initial condition close to the ultraviolet fixed point of QEG.
The tensor track is a proposal in this direction.

\section{Tensor Invariant Interactions}

The tensor theory space is independent of any particular geometry of space and time. 
It generalizes the natural invariance
$U(N)$ of  vector and matrix models (in the complex case), hence
relies on a more abstract symmetry than diffeomorphisms.

The canonical quadratic action of vectors, matrix and tensor models 
is always the same, namely a scalar product. Hence the free part of these models always consist of independent identically distributed Gaussian
distributions for each components. The main difference appears
at the level of interactions and observables, and it can be formulated as a symmetry breaking. This is already true for random matrices: 
rectangular $N_1$ by $N_2$ Wishart matrices are not just rectangular arrays of random numbers,
their polynomial interactions and observables, which are product of traces, have the reduced $U(N_1) \otimes U(N_2)$ symmetry rather than 
a full $U(N_1 \times N_2)$ vector symmetry.

The natural class of interactions and of observables for higher rank tensor models simply corresponds
to further symmetry breaking. Rank $D$ tensors are vectors in a $N_1 \times \cdots \times N_D$-dimensional space, but have only 
a tensorial $U(N_1) \otimes \cdots \otimes U(N_D)$ invariance, which represent 
independent change of orthonormal basis in each of the spaces of the tensor product. The corresponding invariant monomials in the 
tensor coefficients were identified long ago by mathematicians. 
They are labeled by bipartite $D$-regular edge-colored graphs. But surprisingly perhaps\footnote{Tensors have 
no \emph{eigenvalues}; this
may have hampered progress.} they were proposed only recently \cite{Gurau:2011kk,Bonzom:2012hw}  
as the most natural interactions and observables to generalize at higher rank the theory of random matrices.

One can compute the number of connected bipartite $D$-regular edge-colored graphs $Z^c_D (n)$ 
on $n$ vertices \cite{Geloun:2013kta} and $n$ anti-vertices, and for $D \ge 3$ it increases fast with $n$:
\bea
Z^c_1 (n) &=& 1, 0, 0, 0 , 0, ...  \nonumber \\  
Z^c_2 (n) &=& 1, 1,1,1,1, 1, 1, ... \nonumber \\  
Z^c_3 (n) &=& 1, 3, 7, 26, 97, 624,  ... \nonumber \\  
Z^c_4 (n) &=& 1, 7, 41, 604, 13753, ... \nonumber
\eea

Tensor models  with canonical trivial propagator and polynomial invariant interactions 
have a perturbative expansion indexed by Feynman graphs which, under expansion of the internal structure of their bipartite $D$-regular edge-colored vertices,
become bipartite $(D+1)$-regular edge-colored graphs. They typically admit an associated $1/N$ expansion \cite{Gur3} whose leading graphs, called melons \cite{Bonzom:2011zz}, are in one-to one correspondence with $(D+1)$-ary trees. This statement requires the
interactions of these models to be themselves melonic. 

The single \cite{Bonzom:2011zz} and double \cite{GurauSchaeffer,Dartois:2013sra} scaling limits of such models
have now been identified  and correspond to continuous random trees, or branched polymers
\cite{Gurau:2013cbh}Ê. This seems at first sight a step backwards.
Indeed the $1/N$ expansion of vector models is also dominated by trees, whereas
the  $1/N$ expansion of matrix models is dominated by planar maps. 
Branched polymers are certainly not a good approximation to our universe, and planar maps seem closer. 
However tensor invariant interactions and the sub-leading structure
of the $1/N$ tensorial expansion are much richer than their vector and matrix counterpart. 
This is not too surprising, given that tensor models of rank $D$
can effectuate a statistical sum over all manifolds (and many pseudo-manifolds) in dimension $D$ \cite{Gurau:2010nd}.

To go beyond single and double scaling of the simplest models and to find more interesting infrared limits and 
phase transitions probably will require to combine analytic and numerical methods. Tensor field theory,
which we now describe, adds a Laplacian to the propagator of the tensor models to equip them with a full-fledged notion
of renormalization group, which generalizes in a natural way
the matrix renormalization group of the 
Grosse-Wulkenhaar model \cite{Grosse:2004yu}.
Such tensor theories become then suited for a FRG analysis, whose preliminary steps have been 
made in \cite{Eichhorn:2013isa}.


\section{Tensor Field Theories}

Any RG analysis relies on a key technical point, the introduction of \emph{scales}. In the case of random tensors the natural notion of scale is the size
of the tensor spaces; the most symmetric (isotropic or hypercubic) case has $ N_1= \cdots =N_D =N$. 
Although the RG flow is 
ultimately universal, different cutoffs break different invariances of the theory and lead to
very different technical aspects of the computations. The standard formulation of the Wetterich equation uses a momentum cutoff $r$ on the inverse
Laplacian propagator of the theory. Keeping the invariant interactions of tensor models but adding an inverse Laplacian to the propagator
we obtain the enlarged formalism of \emph{tensor field theories} which is therefore suited to FRG. 

Since each Lie group $G$ has a canonical Laplace-Beltrami operator, such a propagator is obtained by
considering tensor theories whose fields $T$ and $\bar T$ are defined on $G^D$,
as in group field theory \cite{Oriti:2013aqa,Krajewski:2012aw}. 
The group also equips tensors with a canonical involution $g \to g^{-1}$
which is interesting in its own right.
We have now a preliminary classification of such theories \cite{Geloun:2013saa,Samary:2012bw}, which  include in particular just renormalizable
models in four dimensions \cite{BenGeloun:2011rc}. Adding a Boulatov-type projector 
to implement the simplicity constraints of the $BF$ theory leads to a related formalism, that of tensorial group field theory  
which includes also super-renormalizable and just renormalizable models with $U(1)$ \cite{Carrozza:2012uv}
or $SU(2)$ groups \cite{Carrozza:2013wda}. It is now established that 
just renormalizable tensor field theories are quite generically asymptotically free 
\cite{BenGeloun:2012pu,BenGeloun:2012yk,Geloun:2012qn,Samary:2013xla}.

In that class of models there should be no particular difficulty to implement FRG truncations. They should 
include increasing combinations of invariants, taking into
account their order (number of vertices), their degree 
(the fundamental integer that governs the tensorial $1/N$ expansion)
and the number of their connected parts. We propose to start by studying at least rank 3 and rank 4 tensors with connected interactions 
up to 6 vertices. Indeed interactions of order 6 tend to be naturally the renormalizable ones in tensor models \cite{BenGeloun:2011rc,Carrozza:2013wda}.
Also remark that order 6 includes the first non planar bipartite graph, $K_{3,3}$, which has no vector or matrix analogue. 
Since the ultraviolet fixed point is Gaussian, the goal is therefore to explore infrared limits, 
which would be a kind of tensorial analogues of confinement in QCD, hence might become analytically difficult at some point. 
We have no doubts that simulation of simple tensor models with melonic interactions
should show the branched polymer phase, but also the planar ``brownian sphere" phase \cite{LeGall}  if properly rescaled
subdominant couplings of the matrix type are added. To go beyond these two phases 
is the real challenge, for which the program of numerical study of tensors through the FRG
is particularly welcome.   

\medskip\noindent
{\bf Acknowledgments}

\medskip

I thank D. Benedetti, A. Eichhorn, R. Gurau, T. Koslowski, M. Reuter, F. Saueressig and R. Sfondrini for useful discussions on the FRG,
and the organizers of the Corfu 2013 summer School for insisting that I should write this note in addition to \cite{Rivasseau:2013uca}.

Research at Perimeter Institute is supported by the Government of Canada through Industry
Canada and by the Province of Ontario through the Ministry of Research and Innovation.

\end{document}